\documentclass{elsart}
\usepackage{graphicx,latexsym,amssymb,natbib}

\newcommand {\sax} {{\it Beppo}SAX }

\newcommand {\al}{\alpha}
\newcommand {\C}{\^Cerenkov }

\newcommand {\gamm} {$\gamma$}
\newcommand {\m}{$\mu$m }
\newcommand {\nufnu}{\nu F_{\nu}}
\newcommand {\nw}{nW/m$^2$/sr }

\begin{document}

\vspace*{-0.1cm}
\begin{frontmatter}
\title{Constraining the Cosmic Background Light with four BL Lac TeV spectra}
\vspace*{-0.35cm}

\author[mpik]{L. Costamante}
\author[mpik]{F. Aharonian}
\author[mpik]{D. Horns}
\author[brera]{G. Ghisellini}

\address[mpik]{Max-Planck Institut f. Kernphysik, Heidelberg, Germany}
\address[brera]{Osservatorio Astronomico di Brera, Merate, Italy}

\vspace*{-0.6cm}
\begin{abstract}
The intrinsic BL Lac spectra above few hundreds GeV can be very different from the observed ones
due to the absorption effects by the diffuse Extragalactic Background Light (EBL), at present poorly known.
With the recent results, there are now 4 sources with good spectral information:
Mkn 421 (z=0.031), Mkn 501 (z=0.034), 1ES 1426+428 (z=0.129) and 1ES 1959+650 (z=0.047).
Making simple assumptions on the shape of the intrinsic spectra (according to the present blazar knowledge),
we have considered the effects of different EBL spectral energy distributions (SED) for the first time 
on all 4 objects together, deriving constraints for the EBL fluxes. 
These resulted significantly lower than many direct estimates. 
\end{abstract}

\end{frontmatter}
\vspace*{-1.2cm}

\section{Introduction}
\vspace*{-0.8cm}
The interpretation of blazars TeV spectra is at present affected by a fundamental ambiguity,
due to the absorption of high energy photons by the diffuse extragalactic background light (EBL) 
through \gamm-\gamm~ collision and pair production. The still large uncertainties
on EBL knowledge prevent a reliable reconstruction of the intrinsic source spectrum,
while conversely data from the synchrotron peak alone are not sufficient to 
univocally constrain the high energy peak of the blazars Spectral Energy Distribution (SED), 
thus hindering the possibility to disentangle absorption from intrinsic features.
Until more and better data are available, 
two possible lines of study are: 1) to assume a ``best estimate" for the EBL
and then to study the reconstructed intrinsic spectrum in the context of blazars emission models 
(see e.g. \cite{co3});
or 2) to make simple, ``reasonable" assumptions on the shapes of the intrinsic TeV spectra 
(according to our present blazar knowledge), 
and then to limit the range of possible EBL fluxes (and spectral shapes) compatible with them.   

This second approach becomes more constraining as more sources are available, since the EBL
must be the same for all and how much an incident spectrum is deformed depends strongly
on redshift.
With the recent results on 1ES 1426+428 \cite{aha3,dj,petry}
and 1ES 1959+650 \cite{horn19,bruno}, we have now 4 objects with measured TeV spectra
(see Fig. 1) and at 3 very different redshifts ($\sim$0.03, 0.047, 0.129), thus allowing
to draw multiple constraints.
In the following, we will 
focus on this second approach, and present the first preliminary results on the EBL SED
obtained considering the effects on all 4 TeV sources at the same time.

\begin{figure*}[t]
\centering
\vspace*{-1.0cm}
\resizebox{15cm}{!}{\includegraphics[angle=0, width=14cm]{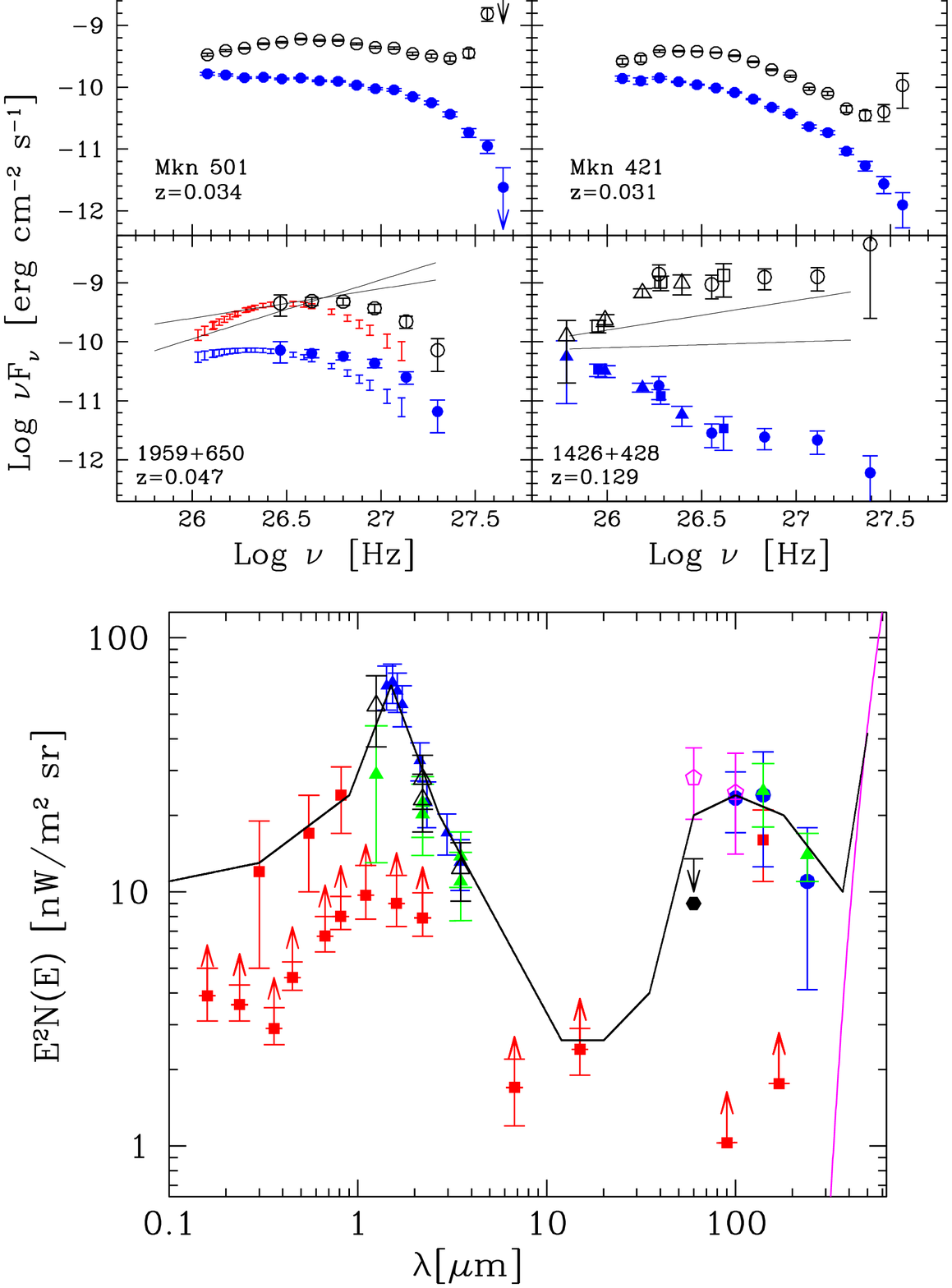} 
\hspace*{0cm} \includegraphics[angle=0, width=14cm]{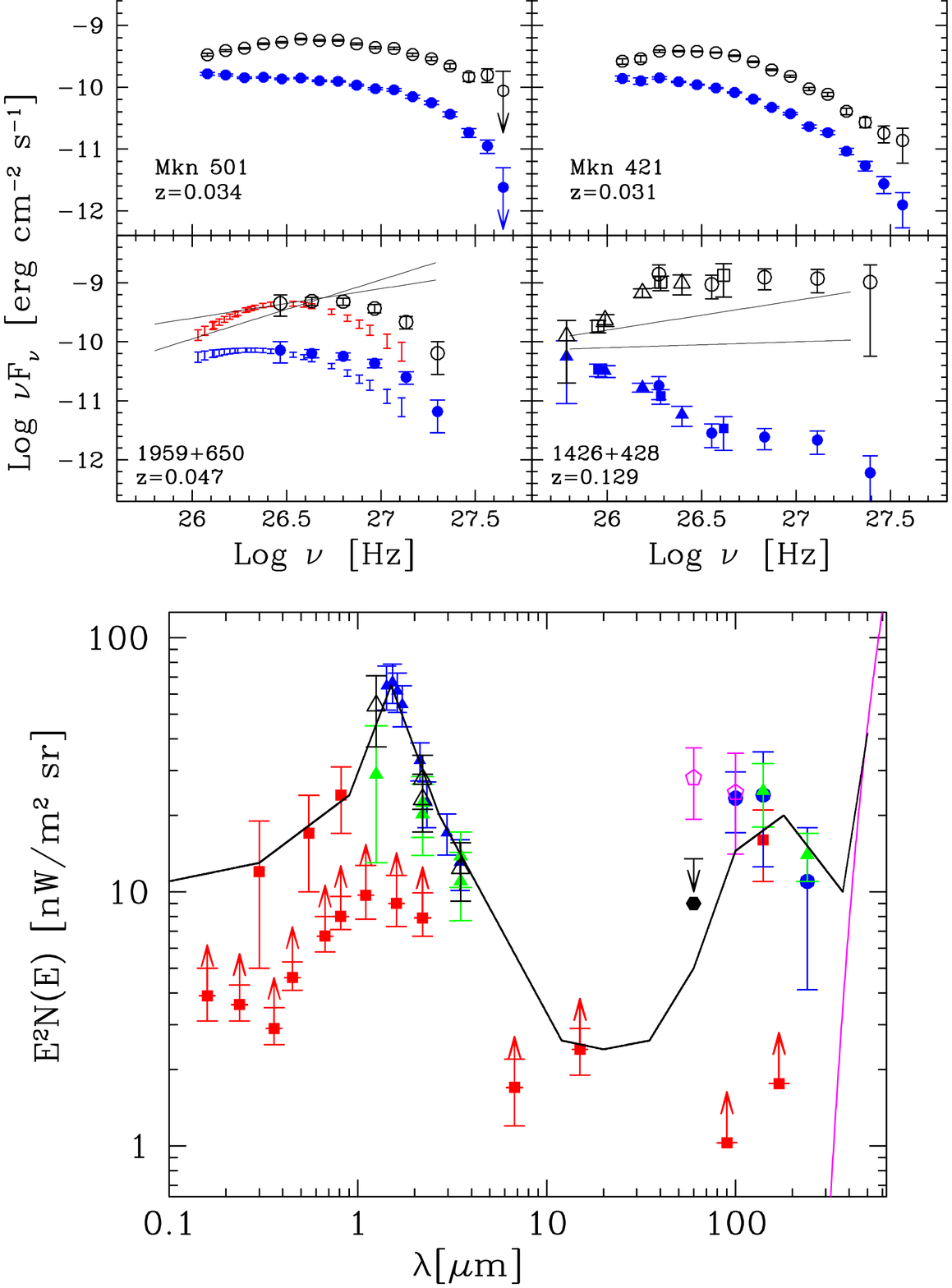} }
\vspace*{-1.2cm}
\caption{\footnotesize Lower panels: EBL data points and the overall shape (continuous line)
used to calculate the absorption-corrected TeV spectra showed in the corresponding upper panels.
Upper panels: filled symbols, the observed data; open symbols, the absorption corrected values.
Circles: HEGRA data; triangles: CAT data; squares: WHIPPLE data. For 1ES 1959+650, the CAT results
are presented as the shape of the fitting model (a log-parabolic curve; details in \cite{bruno}); 
the CAT and HEGRA data correspond to 
the high states in May-June and May$+$July 2002, respectively.
As reference for the eye, in the 1ES 1959+650 and 1ES 1426+428 subpanels, 
two power-law spectra are also shown, with energy index $\al=0.0$ and 0.5, and 0.5 and 0.9 respectively
($F_{\nu}\propto \nu^{-\alpha}$).
Left: EBL shape matching the reported measurements. Right: as left, but with lower fluxes in 
the 30-100 \m range  
to avoid  Mkn 501 and Mkn 421 high energy upturns 
(as a reminder, the threshold condition for \gamm-\gamm~ absorption is given by  
$\lambda\leq\lambda_{thr}\backsimeq4.75(E/1{\rm TeV})$ \m). }
\label{f1} 
\end{figure*}

\vspace*{-1.0cm}
\section{Data and Results}
\vspace*{-0.8cm}
Fig. 1 and 2 show  different EBL shapes (obtained simply by connecting flux values at fixed frequencies)
and the corresponding effects on the absorption--corrected TeV spectra. 
The compilation of measurements and lower limits have been taken from \cite{haus}, updated with the most 
recent publications (ref. in \cite{aha3}).
For the blazar spectra, we have used the HEGRA data \cite{aha1,aha2,horn19,aha3}
completed at low energy by the CAT and WHIPPLE results for 1ES 1426+428 \cite{dj,petry}
and 1ES 1959+650 \cite{bruno}.
In calculating the optical depths, we have assumed 
no model for EBL evolution, but only the changes in energy and photon density due to cosmology.
For the latter, we adopted the recent W-MAP results, namely:
H$_0=70$ km/s/Mpc, $\Omega_{mat}=0.3$ and $\Omega_{\Lambda}=0.7$.

\begin{figure*}[t]
\centering
\vspace*{-1.0cm}
\resizebox{15cm}{!}{\includegraphics[angle=0, width=14cm]{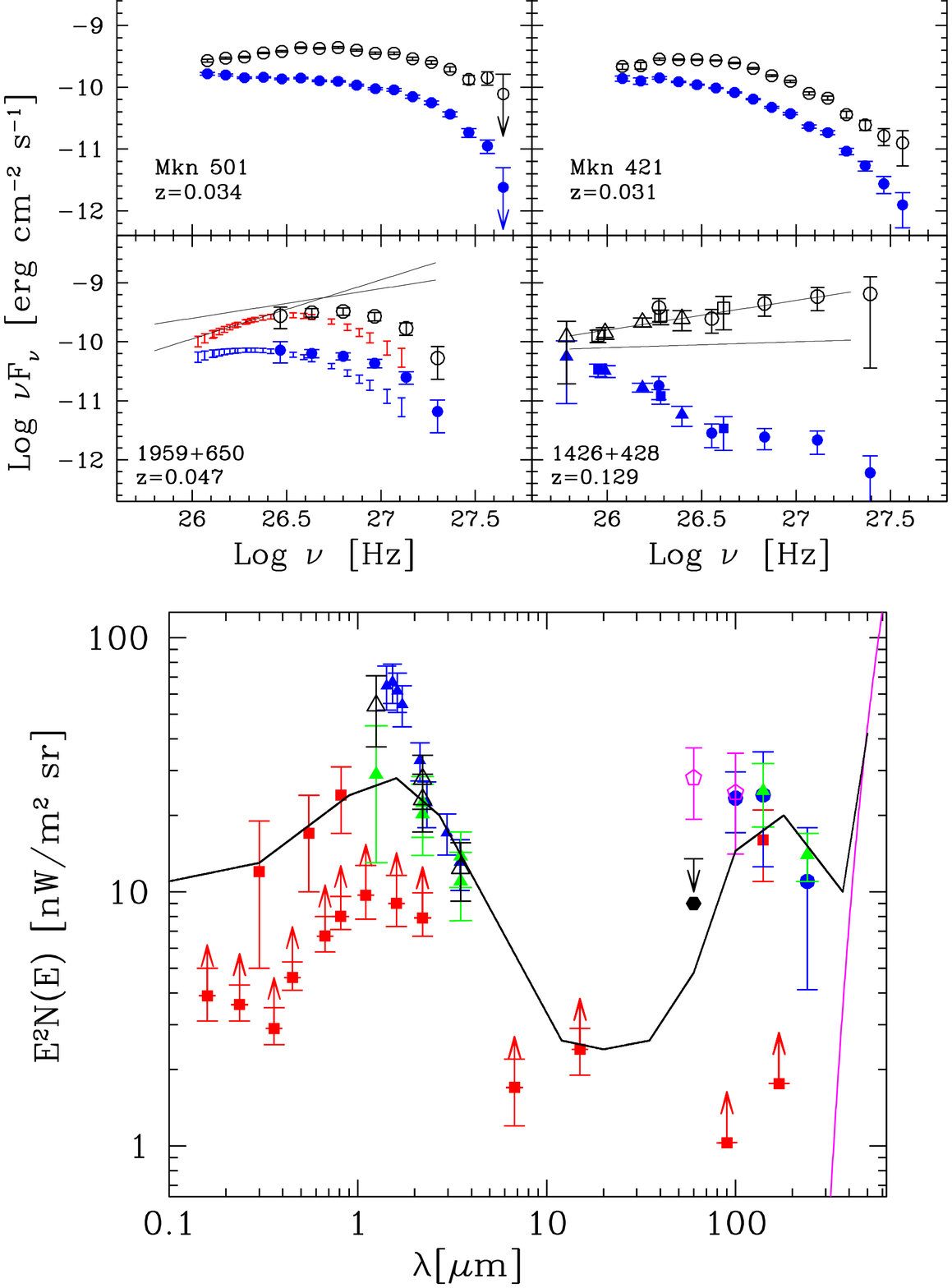} 
\hspace*{0cm} \includegraphics[angle=0, width=14cm]{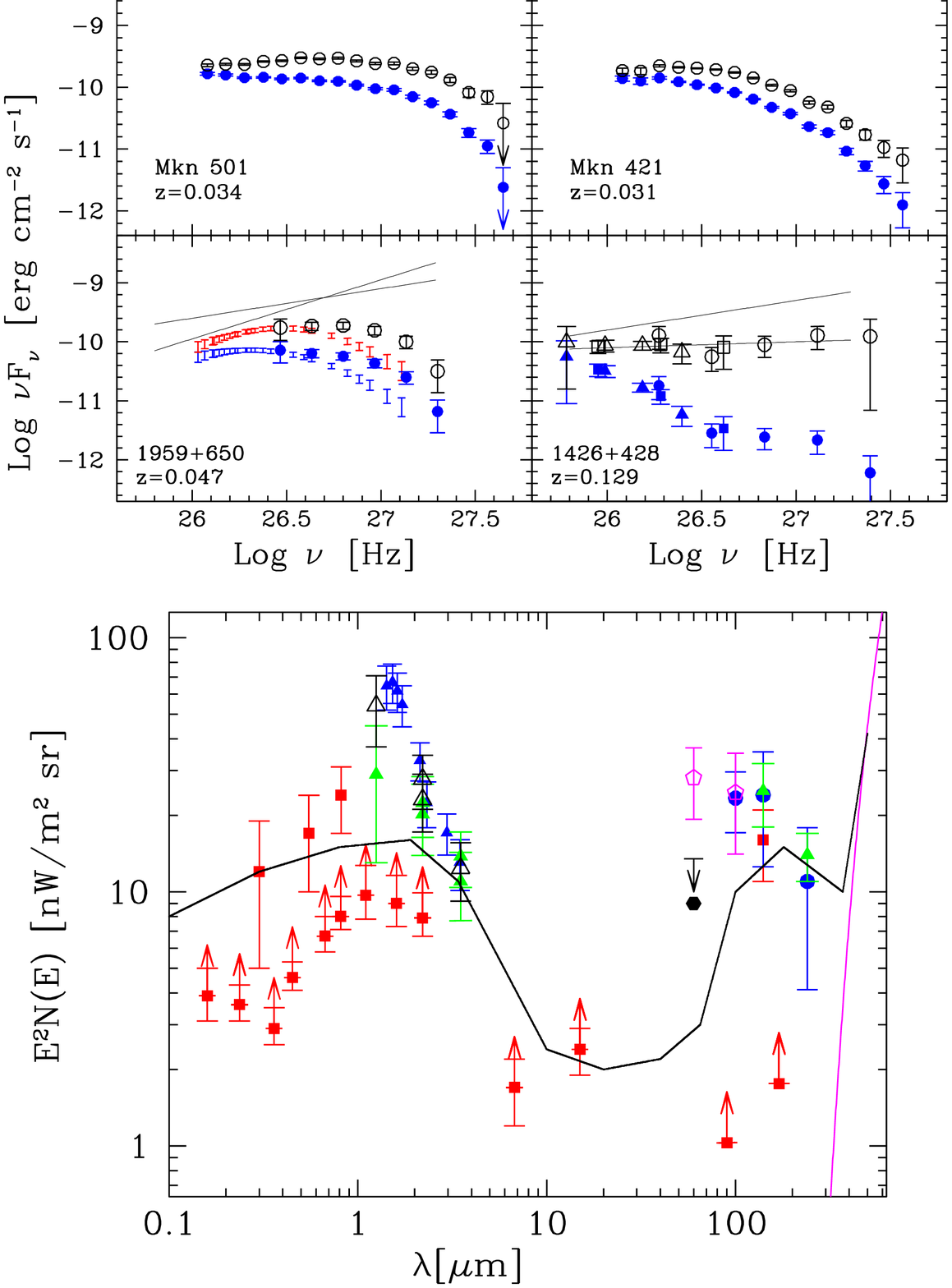} }
\vspace*{-1.2cm}
\caption{\footnotesize Left: as Fig. 1 right, but with lower fluxes at 1-3 \m to avoid 
strange spectral shapes and spectra harder than $\al=0$. Right: further changing the overall shape
to obtain an intrinsic 1ES 1959+650 spectrum between 0.5 and 1 TeV not harder than $\al\sim0.5$,
and the 1ES 1426+428 spectrum similar to the \sax X-ray one ($\al\sim0.9$).}
\label{f2}
\end{figure*}

Our main assumptions for a ``reasonable" blazar TeV spectrum have been: a smooth 
(power-law or curved) continuum, with no sharp features as up-turns or strong breaks,
and a spectral index $\alpha$ ($F_{\nu}\propto \nu^{-\alpha}$) not too hard, i.e. possibly $\geq 0.5$,
which is the hardest value expected from an electron distribution cooled by synchrotron or inverse Compton
mechanisms. 
Fig. 1 (left) show the results with an EBL shape designed to match the reported EBL measurements, in
particular the high points at 1-3 \m \cite{mats}, and the first estimates at 60 \m \cite{fink}.
This gives origin to a sharp up-turn in the Mkn 501 and Mkn 421 spectra, and to an extremely 
hard spectrum up to $\sim$1 TeV for 1ES 1959+650 and 1ES 1426+428 
($\al\approx-1$, i.e. $\nufnu\propto\nu^2$), with a sharp 
break for the latter. To avoid these effects, it is necessary to lower the EBL flux
in the range 30-80 \m (to avoid the up-turns in the two Markarians, Fig. 1 right) and in the range
1-3 \m (to reduce the hardening of the low energy spectra, Fig. 2 left).
Note also that the slope between 2 and 10 \m should be around (or steeper than) 
$\lambda F_{\lambda}\propto\lambda^{-1}$, 
otherwise the flattening of the observed 1ES 1426+428 spectrum above 2 TeV would imply
again a sharp up-turn and very hard slope for the intrinsic one \cite{horn,aha3}.

With an EBL SED given in Fig. 2 (left), which is similar to some model predictions \cite{primack},
one can avoid strange shapes for the blazar intrinsic spectra,
but still has to face two uncommon features: a very hard spectrum for 1ES 1959+650
($\alpha\approx0$ between 0.5 and $\sim$2 TeV) and the overall SED of 1ES 1426+428, whose TeV spectrum
would imply a high energy peak at or above $\sim$8 TeV with a Compton dominance $L_C/L_S\geq10$, 
unexpected in
this class of objects \cite{co3}. To avoid this, the EBL fluxes should be of even lower
intensity below $\sim$3 \m, approaching the lower limits given by the HST counts
\cite{madau}.
With a shape as given in Fig. 2 (right), 
the intrinsic 1ES 1959+650 spectrum would have a low energy slope around $\al\sim$0.5, and
1ES 1426+428 would be characterized by a lower (more usual) Compton dominance ($L_C$$\sim$$L_S$)
with a TeV spectral slope similar to the X-ray one ($\alpha$$\sim$0.9, \cite{co,horn}).   
These limits can be loosened a bit taking into account 
the residual uncertainty on the absolute flux and energy calibration of \C telescopes, 
estimated in $\sim$20\% and $\sim$15\% respectively
(i.e. systematically decreasing energy and normalization of the data points of all the 4 TeV sources).

\vspace*{-1cm}
\section{Conclusions}
\vspace*{-0.8cm}
To avoid ``unreasonable" shapes for the blazar intrinsic spectra,
at least according to our present knowledge of the blazar phenomenon,
the data of the 4 TeV BL Lacs considered together put some limits to the EBL fluxes and spectral
shape, which can be roughly 
given as:  $\lesssim$15 \nw in the range 0.8-3 \m,  $\lesssim$6 \nw at 
60\m (these values become $\lesssim$20 and $\lesssim$10 nW/m$^2$/sr, respectively, 
including the systematic error),
and with a slope around or steeper than $\lambda F_{\lambda}\propto\lambda^{-1}$ between 2 and 10 \m. 
Conversely, if higher values or different shapes should be confirmed by more constraining data or models, 
this would require a deep revision of our understanding of the blazars emission mechanisms and physical
conditions.


\end{document}